\documentclass{aa}
\usepackage{graphicx,psfig,amssymb,natbib}

\def\hunits{$\rm km \ s^{-1} Mpc^{-1}$}
\def\nh{$N_{\rm H}$} 
\def\lunits{erg~s$^{-1}$}
\def\lxo{$\log (f_X/f_O)$}
\def\lxol2{$\log (f_X/f_O) < -2$}
\def\ran{$-2 < \log (f_X/f_O) < -1$}
\def\funits{erg~cm$^{-2}$~s$^{-1}$}

\def\xmm{{\it XMM-Newton}}
\def\chandra{{\it Chandra}}

\def\2df{{\it 2dFGRS}}

\def\ha{H$\alpha$}
\def\hb{H$\beta$}
\def\hone{\ion{H}{i}}
\def\htwo{\ion{H}{ii}}
\def\kmps{km s$^{-1}$}
\def\ntwof{[\ion{N}{ii}]}
\def\othreef{[\ion{O}{iii}]}

\def\oonef{[\ion{O}{i}]}
\def\stwof{[\ion{S}{ii}]}

\begin{document}

\title{Searching for X-ray luminous \lq normal\rq\ galaxies in \2df} 

\author{P.~Tzanavaris\inst{1} \and  I.~Georgantopoulos\inst{1}
\and  A. Georgakakis\inst{2}}
\offprints{P.~Tzanavaris; email: pana@astro.noa.gr}

\institute{Institute of Astronomy \& Astrophysics, \ National
Observatory of Athens, I.~Metaxa \& V.~Pavlou, Penteli 15236, Greece
\and Astrophysics Group, Imperial College London, Blackett Laboratory,
Prince Consort Rd, SW7 2AW, United Kingdom} \date{Received... ;
accepted... }  \abstract{We cross-correlated the Chandra XASSIST and
XMM-Newton Serendipitous Source Catalogues with the 2 degree Field Galaxy
Redshift Survey (\2df) database.  Our aim was to identify the most
X-ray luminous ($L_X > 10^{42}$~\lunits) examples of galaxies in the
local Universe whose X-ray emission is dominated by stellar processes
rather than AGN activity (\lq normal\rq\ galaxies) as well as to test
the empirical criterion \lxo~$< -2$ for separating AGN from NGs.  With
\xmm\ (\chandra) we covered an area of $\sim 8.2$ ($\sim
5.8$)~deg$^2$ down to a flux limit of $\sim 10^{-15}$
($\sim 1.6 \times 10^{-15}$)~\funits\ and found 18 (20) \2df\ galaxies.  Using
emission-line intensity ratios, we classified 6 \2df\ spectra as
star-forming, \htwo\ nuclei, and 2 spectra as possible \htwo\
nuclei. The rest of the objects are absorption-line galaxies and AGN,
including 3 possible LINERs.  No luminous \lq normal\rq\ galaxies
have been found but out of 19 \lq normal\rq\ galaxies in this sample 
5 \htwo\ and 3 absorption-line galaxies have
\lxo~$> -2$. We performed a similar search in two nearby-galaxy
samples from the literature. All 44 galaxies in the \citet{ZezasPhD} sample
have \lxol2\ and $L_X < 10^{42}$~\lunits. In the \citet{1992ApJS...80..531F} sample,
out of a total of 170 \lq normal\rq\ galaxies,
we found 16
galaxies with \lxo~$>-2$, the majority of which are massive
ellipticals. Three of these have $L_X > 10^{42}$~\lunits.

 }

\titlerunning{X-ray luminous \lq normal\rq\ galaxies in \2df }
\authorrunning{P.~Tzanavaris et al.}  

\maketitle

\section{Introduction}
Galaxies hosting an active galactic nucleus (AGN) have long been known
to be sources of copious amounts of X-ray emission. On the other hand, galaxies
which are \lq normal\rq\ (NGs), in the sense that they are not AGN
dominated, have also
been studied in detail. The source of X-ray emission in these
systems is diffuse hot gas and/or X-ray binary stars. In the most
massive early-type galaxies the X-ray emission is dominated by the hot
interstellar medium with temperatures $kT\sim 1$~keV. A smaller
fraction of the observed X-ray luminosity is due to low mass X-ray
binaries associated with the older stellar population. In late-type
galaxies, the X-ray emission originates in hot gas with temperature
$kT\sim 1$~keV, which is heated by supernova remnants, as well as in a
mixture of low and high-mass X-ray binaries \citep[see][for a
review]{1989ARA&A..27...87F}. The diffuse hot gas contributes
significantly in the soft X-ray band ($<2$~keV) while the X-ray binary
systems are responsible for the bulk of the emission at harder
energies \citep[e.g.][]{2003MNRAS.343L..47S}.  The integrated X-ray
emission of \lq normal\rq\ galaxies is believed to be a good indicator of
the star-formation activity in these
systems \citep[e.g.][]{2004MNRAS.347L..57G}, at least if
the star formation rate is not too low.

The X-ray luminosity of NGs is usually weak, $\la 10^{42}$~\lunits,
i.e. a few orders of magnitude below that of powerful AGN
\citep{1999ApJ...526..649M,1998MNRAS.301..915Z}. As a result, observed
X-ray fluxes are faint and, until recently, only the very local
systems ($\rm <100 \, Mpc$) were accessible to X-ray missions. With
the new generation of X-ray missions, \chandra\ and \xmm, the
situation has changed dramatically. The \chandra\ Deep Fields North
and South
\citep[CDF-N, CDF-S;][]{2003AJ....126..539A,2002ApJS..139..369G} have reached
fluxes $f(0.5-2.0 \ {\rm keV}) \sim 10^{-17}$~\funits, thus providing
the first ever X-ray selected sample of NGs at cosmologically
interesting redshifts.  Using the 2\,Ms CDF-North,
\citet{2003AJ....126..575H} provided a sample of 43 NG candidates for
which optical spectroscopic observations are
available. These galaxies have X-ray--to--optical flux ratios 
\lxo~$\la -2$, which these authors use as an empirical boundary,
separating quiescent NGs from AGN.
\citet{2004ApJ...607..721N} extended this study and
identified over 100 NG candidates in the combined CDF-N and CDF-S,
although optical spectroscopic data are available only for a fraction
of these objects. However, these authors have included NGs
with \lxo~$>-2$. On the other hand, \citet{2003MNRAS.344..161G}
and \citet{2004MNRAS.349..135G} have identified NGs with
\lxo~$\approx -2$. Within the framework of the \lq Needles in the
Haystack Survey\rq\ (NHS) \citet{2004MNRAS.354..123G} and
\citet{2005MNRAS.360..782G}
combined \xmm\ data with the Sloan Digital Sky Survey 
and used several selection criteria, including 
\lxo~$< -2$, to identify 28 NG candidates. By combining this
sample with 18 $z<0.2$ galaxies from the CDFs these authors
constructed the first local X-ray luminosity function of
NGs. However, this result depends crucially on
the completeness of NG samples, which, in turn,
may be biasing to quiescent systems and selecting
against X-ray luminous (with X-ray luminosities $L_X>10^{42}$\lunits)
starbursts and massive ellipticals, which are likely to
show \ran\ \citep{2003AJ....126..539A}. 
Indeed, the local luminosity function of
\citet{2005MNRAS.360..782G} agrees well with that of 
\citet{2004ApJ...607..721N} at the faint end but 
disagrees at the bright end. If bright galaxies
are missed due to the \lxol2\ criterion, this might
explain the discrepancy.
Alternatively, the \citet{2004ApJ...607..721N}
may suffer from AGN contamination.

It is thus imperative to understand the significance of this bias and
to resolve the controversy in order to constrain the local luminosity
function, which also provides a local \lq anchor point\rq\ for
investigating luminosity function evolution.

In the work described in this paper we searched for 
luminous NGs by performing a cross-correlation between, on
the one hand, two large X-ray catalogues, and, on the other hand, the
2 degree field galaxy redshift survey (\2df). 
Our aim was twofold:
\begin{enumerate}
\item to search for X-ray luminous ($L_X > 10^{42}$~\lunits),
especially star-forming galaxies, and
\item to test the empirical criterion \lxo~$< -2$ for separating
NGs from AGN so as to address the discrepancy described above.
\end{enumerate}
Throughout this paper we use a Hubble parameter $H_0=72$~\hunits,
matter density $\Omega_{\rm M}=0.3$ and a cosmological
constant $\Omega_{\Lambda}=0.7$.
\section{Data}

\subsection{2dfGRS data}
The \2df\ is a joint UK-Australian project, which obtained spectra for
245,591 objects brighter than a nominal extinction-corrected magnitude
limit of $b_J=19.45$ over an area of $\approx 1500 \ {\rm
deg}^2$. This survey is fairly uniform with a known incompleteness at
the bright end \citep[$b_J \la 16$,][]{2002MNRAS.336..907N}. Reliable
redshifts were obtained with the aim of providing a detailed
three-dimensional picture of galaxy population and large scale
structure in the local Universe
\citep{2003Colless,2001MNRAS.328.1039C}. We used heliocentric
corrected redshift values given in the FITS headers of the \2df\
spectra.  The galaxies are selected from the extended APM Galaxy
Survey in three regions: a strip near the north galactic pole, a strip
near the south galactic pole and random fields around this.

For our purposes the \2df\ presents the advantage that its depth
allows detection of galaxies up to \lxo~$ \la -1$ for $f_X \approx \
{\rm few} \times 10^{-14} $ (see Fig.~\ref{fig:bjfx}).

\subsection{XMM-Newton observations}\label{sec:xmm}
We used \xmm\ archival observations from the XMM-Newton Serendipitous
Source Catalogue, Version 1.1.0, (1XMM), whose fields overlap with
those from the final data release of the \2df. The catalogue contains
more than 50,000 single sources in a total of 585
fields. Specifically, we have cross-correlated the celestial
coordinates (right ascension and declination, J2000) of sources in the
two data sets in order to detect optical and X-ray counterparts.  The
cross-correlation was performed independently for each of the three
CCDs of the EPIC camera because in some cases a source may not have
been recorded in all three CCDs. The X-ray band was restricted to $0.5
- 2.0$~keV as galaxies are preferentially soft X-ray emitters
\citep[e.g.][]{2001ApJ...550..230L}. We set the detection likelihood
parameter to $\ga 7$ and the matching radius to 6\arcsec. This
corresponds to $\sim 3\sigma$ in terms of on-axis XMM/EPIC positional
uncertainty. Off-axis sources suffer from vignetting and such sources
will have increased positional uncertainty. This is further increased
for sources whose X-ray emission is dominated by an off-nuclear
component.

We obtained 18 sources which have been detected both by \2df\ and by
XMM/EPIC. The sources are detected in 42 \xmm\
fields. The area covered for this sample is $\sim 8.2 {\rm deg}^2$
to a flux limit of $\sim 10^{-15}$~\funits.
The largest separation between an X-ray and an optical
position in this sample is 5.1\arcsec. All X-ray/optical counterparts
were checked visually.
For all sources we estimate the probability of detecting
an optical
counterpart by chance to be less than 1\% for all sources.  
These 18 X-ray/optical pairs form our \xmm--\2df\ correlation sample.

We calculated X-ray fluxes, $f_X$, after taking into account the
column density of Galactic neutral hydrogen, \nh, along the line of
sight to each observed source.  Further, we calculated the X-ray
luminosity, $L_X$, by using source redshifts, $z$, and X-ray fluxes,
assuming power law spectra with a photon index $\Gamma=1.8$.

We obtained hardness ratio values, HR, from the 1XMM catalogue, using
unvignetted count rates in the energy bands 0.5-2.0~keV (S) and
2.0-4.5~keV (H), so that
\begin{equation}\label{equ:HR}
{\rm HR}\equiv \frac{\rm H-S}{\rm H+S} \ .
\end{equation}

Details of our \xmm--\2df\ correlation sample are given in Table~\ref{tab:xmm}.

\begin{table*} 
\scriptsize
\caption{The \xmm/\2df\ correlation sample. Shown from left to right are sample
identification number, \2df\ database name, right ascension and declination for
the X-ray source, offset between X-ray and optical source, $b_J$ magnitude from
\2df, X-ray flux, redshift from \2df, logarithm of X-ray luminosity, 
logarithmic X-ray-to-optical flux ratio, hardness ratio and error, and, in
the last column, suggested galaxy type. For this column the following abbreviations
hold: A: absorption line galaxy; F: featureless \2df\ spectrum; G: source appears in group; \htwo: \htwo\
nucleus. A question mark indicates
that a classification is not possible. Question marks after a
classification indicate a high degree of uncertainty.} 
\label{tab:xmm} 
\centering 
\begin{tabular}{cccc cccc cccc l} 
\hline\hline 
ID &       Name    & $\alpha_X$     & $\delta_X$ & $\delta_{XO}$ &$b_J$& $f_X/10^{-14}$     & $z$   & log $L_X$& \lxo\ & HR & $\pm$ & Type \\ 
   &               & (J2000)        & (J2000)    &   (\arcsec)   &     &(cgs)& & (cgs)&      &   &       &          \\
\hline 
1  & 	\object{TGS548Z244} &	23 56 27.68 & $-34$ 35 35.8 & 0.65 & 	15.52 & 1.700 & 0.0479 & 40.94 & $-2.09$ & $-1.00$ & 0.11 & A \\
2  & 	\object{TGS549Z357} &	23 56 10.72 & $-34$ 49 42.1 & 3.20 & 	18.73 & 1.170 & 0.2399 & 42.26 & $-0.97$ & $-0.74$ & 0.16 & F, G\\
3  & 	\object{TGS617Z146} &	00 58 18.33 & $-35$ 55 48.1 & 3.12 & 	18.01 & 0.375 & 0.0479 & 40.28 & $-1.75$ & $-0.57$ & 0.15 & \htwo\ \\
4  & 	\object{TGS210Z018} &	00 55 51.51 & $-27$ 26 09.7 & 3.19 & 	19.12 & 4.555 & 0.2125 & 42.74 & $-0.22$ & $-0.50$ & 0.24 & F\\
5  & 	\object{TGS327Z003} &	21 51 06.12 & $-30$ 24 27.1 & 1.72 & 	18.83 & 2.080 & 0.1373 & 42.00 & $-0.68$ & $-0.77$ & 0.17 & A, G \\
6  & 	\object{TGS063Z133} &	22 21 51.26 & $-24$ 45 30.7 & 0.98 & 	19.08 & 2.470 & 0.0777 & 41.54 & $-0.51$ & $-0.26$ & 0.09 & AGN\\
7  & 	\object{TGS120Z043} &	22 35 32.80 & $-26$ 06 05.6 & 3.83 & 	15.37 & 0.321 & 0.0192 & 39.40 & $-2.87$ & $-0.62$ & 0.15 & \htwo\ \\
8  & 	\object{TGN296Z199} &	10 44 44.83 & $-01$ 20 17.7 & 1.35 & 	18.73 & 2.230 & 0.1846 & 42.29 & $-0.69$ & $-0.71$ & 0.04 & ? \\
9  & 	\object{TGN295Z042} &	10 44 19.91 & $-01$ 24 26.7 & 1.97 & 	18.69 & 0.158 & 0.0610 & 40.13 & $-1.86$ & $-0.37$ & 0.17 & \htwo\ \\
10  &	\object{TGN295Z067} &	10 43 52.59 & $-01$ 17 40.1 & 2.97 & 	14.69 & 3.330 & 0.0262 & 40.70 & $-2.13$ & $-0.75$ & 0.03 & \htwo\ \\
11  &	\object{TGN448Z020} &	11 51 29.69 & $+01$ 48 31.7 & 5.81 & 	18.80 & 0.211 & 0.1581 & 41.13 & $-1.68$ & $-0.80$ & 0.17 & F \\
12  &	\object{TGN388Z113} &	12 29 47.33 & $+01$ 54 03.5 & 1.03 & 	19.04 & 1.360 & 0.1577 & 41.94 & $-0.78$ & $-0.77$ & 0.11 & ? \\
13  &	\object{TGN387Z032} &	12 28 58.43 & $+02$ 11 27.2 & 0.14 & 	16.64 & 0.500 & 0.0775 & 40.84 & $-2.18$ & $-1.00$ & 0.07 & A\\
14  &	\object{TGN387Z056} &	12 28 17.84 & $+02$ 12 29.3 & 0.55 & 	18.86 & 3.850 & 0.1203 & 42.16 & $-0.40$ & $-0.66$ & 0.11 & ?\\
15  &	\object{TGN387Z067} &	12 28 07.61 & $+02$ 02 52.1 & 0.66 & 	18.48 & 0.694 & 0.0903 & 41.12 & $-1.30$ & $-0.61$ & 0.19 & ? \\
16  &	\object{TGN266Z089} &	13 31 39.59 & $-01$ 48 26.2 & 4.79 & 	18.23 & 0.290 & 0.0749 & 40.57 & $-1.77$ & $-0.26$ & 0.20 & \htwo\ \\
17  &	\object{TGS924Z253} &	22 52 28.19 & $-17$ 49 16.8 & 0.78 & 	19.24 & 3.290 & 0.1331 & 42.16 & $-0.31$ & $-0.63$ & 0.05 & ? \\
18  &	\object{TGS924Z246} &	22 52 31.62 & $-17$ 46 32.4 & 0.60 & 	18.44 & 0.853 & 0.0689 & 40.97 & $-1.22$ & $-0.43$ & 0.23 & \htwo\ \\
\hline 
\end{tabular} 
\end{table*}

\begin{table*} 
\scriptsize
\caption{The \chandra/\2df\ correlation sample. Column details
are as in the previous figure.} 
\label{tab:chandra} 
\centering 
\begin{tabular}{cccc cccc cccc l} 
\hline\hline 
ID &       Name    & $\alpha_X$     & $\delta_X$ & $\delta_{XO}$ &$b_J$& $f_X/10^{-14}$     & $z$   & log $L_X$& \lxo\ & HR & $\pm$ & Type \\ 
   &               & (J2000)        & (J2000)    &   (\arcsec)   &     &(cgs)& & (cgs)&      &   &       &          \\
\hline 
 1 & 	\object{TGS432Z052} & 	23 58 27.21 & $-32$ 41 03.3 & 0.23         & 	19.20	  & 6.36 & 0.2394 &  42.99 & $-0.50$ &  +0.39 &  0.04  &  AGN$^{1}$   \\
 2 & 	\object{TGS522Z150} & 	02 57 14.04 & $-33$ 18 27.5 & 0.82         & 	17.93	  & 1.84 & 0.1093 & 41.72 & $-1.54$ &  +0.23 &    0.11 &  AGN$^{1}$  \\
 3 & 	\object{TGS212Z026} & 	00 58 24.51 & $-27$ 29 23.2 & 0.78         & 	16.83	  & 0.36 & 0.0977 & 40.91 & $-2.70$ & $-0.63$ &   0.24 &  \htwo?    \\
 4 & 	\object{TGS211Z082} & 	00 56 51.78 & $-27$ 28 56.3 & 1.44         & 	18.90	  & 4.22 & 0.2144 & 42.71 & $-0.80$ &  +0.61 &    0.11 &  AGN$^{1}$   \\
 5 & 	\object{TGS243Z005} & 	03 32 46.89 & $-27$ 42 13.3 & 2.52         & 	17.53	  & 0.39 & 0.1028 & 41.00 & $-2.38$ & $-0.54$ &   0.12 &  A        \\
 6 & 	\object{TGS407Z114} & 	22 01 36.09 & $-31$ 53 23.2 & 1.63         & 	18.06	  & 1.63 & 0.0972 & 41.56 & $-1.55$ &  +0.24 &  0.03 &  AGN$^{1}$ \\
 7 & 	\object{TGS132Z150} & 	23 51 39.38 & $-26$ 05 02.5 & 2.90         & 	18.00	  & 9.17 & 0.2346 & 43.13 & $-0.82$ & $-0.61$ &  0.01 &  LINER? \\
 8 & 	\object{TGN163Z121} & 	10 56 50.04 & $-03$ 33 42.8 & 1.10         & 	17.69	  & 0.69 & 0.0488 &   40.56 & $-2.07$ & $-0.56$ &  0.07 &  A (noisy)   \\
 9 & 	\object{TGN163Z123} & 	10 56 48.84 & $-03$ 37 25.8 & 0.73         & 	18.54	  & 0.41 & 0.1816 & 41.55 & $-1.96$ & $-0.78$ &   0.10 &  A   \\
10 & 	\object{TGN071Z177} & 	12 49 02.23 & $-05$ 49 33.8 & 1.43         & 	18.34	 & 0.60 & 0.0485 & 40.50 & $-1.87$ & $-0.32$ &  0.08 &  \htwo?  \\
11 & 	\object{TGN206Z127} & 	14 12 49.69 & $-03$ 07 19.7 & 2.59         & 	18.09	 & 0.94 & 0.0748 & 41.08 & $-1.77$ & $-0.43$ &   0.13 &  AGN? \\
12 & 	\object{TGN440Z057} & 	11 23 20.17 & $+01$ 38 09.9 & 1.84         & 	18.66	 & 0.07 & 0.1259 & 40.47 & $-2.66$ & $-0.68$ &   0.25 &  A   \\
13 & 	\object{TGN382Z123} & 	12 04 27.15 & $+01$ 53 46.2 & 2.69         & 	14.25	 & 3.10 & 0.0197 & 40.42 & $-2.79$ & $-0.80$ & 0.01 &  A   \\
14 & 	\object{TGN243Z240} & 	11 55 45.38 & $-01$ 41 30.3 & 0.70         & 	19.00	 & 1.79 & 0.2476 & 42.48 & $-1.13$ & $-0.57$ &  0.08  &  LINER?   \\
15 & 	\object{TGN247Z076} & 	12 16 01.62 & $-00$ 37 33.6 & 0.46         & 	17.70	 & 0.25 & 0.1214 & 40.98 & $-2.50$ & $-1.00$ &   0.24 &  A \\
16 & 	\object{TGN318Z209} & 	12 15 53.29 & $-00$ 36 06.9 & 1.71         & 	18.00	 & 0.23 & 0.1193 & 40.90 & $-2.42$ & $-0.39$ &   0.21 &  A  \\
17 & 	\object{TGN336Z169} & 	13 44 52.88 & $+00$ 05 20.4 & 0.81         &      17.38	 & 77.8 & 0.0876 & 43.14 & $-0.14$ & $-0.63$ &  0.02 &  AGN?$^{2}$ \\
18 & 	\object{TGN336Z187} & 	13 44 28.34 & $+00$ 01 47.2 & 0.71         & 	18.12	 & 0.56 & 0.1351 & 41.43 & $-1.98$ & $-1.00$ &   0.48 &  A  \\
19 & 	\object{TGN275Z203} & 	14 12 34.68 & $-00$ 35 00.1 & 0.49         & 	18.16	 & 30.4 & 0.1269 & 43.09 & $-0.24$ & $-0.74$ &  0.07 &  AGN?$^{3}$  \\
20 & 	\object{TGS906Z501} & 	23 25 19.74 & $-12$ 07 26.2 & 1.77         & 	15.89	 & 4.94 & 0.0824 & 41.89 & $-1.93$ & $-0.64$ & 0.002 &  LINER? \\
\hline 

\multicolumn{13}{l}{1: Obscured AGN suggested by high \nh\ value.}\\
\multicolumn{13}{l}{2: Suggested by a an \hb\ FWHM $\sim 2000$\kmps.}\\
\multicolumn{13}{l}{3: Suggested by a an \hb\ FWHM $\sim 1500$\kmps.}\\

\end{tabular} 
\end{table*}

\subsection{Chandra observations}
We used \chandra\ archival observations from the \chandra\ XAssist
catalogue (version 3).  XAssist
\citep[http://www.xassist.org,][]{2003ASPC..295..465P} is a NASA
AISR-funded project (NAG 5-809) for the automation of X-ray
astrophysics, with emphasis on galaxies. The XAssist catalogue
comprises more than 40,000 single X-ray sources detected in the $0.3 -
8.0$ keV band in a total of 913 fields. We cross-correlated celestial
coordinates in the XAssist and \2df\ catalogues for overlapping fields
from the two databases by using a procedure similar to that used to
obtain the \xmm-\2df\ correlation sample. We used a matching radius of
3\arcsec\ which corresponds to $\ga 3\sigma$ in terms of on-axis
positional uncertainty for ACIS.
As in the case of \xmm, we also need to account for a large increase
in this uncertainty for off-axis and off-nuclear sources. 

We obtained 20 sources for which the optical and X-ray sources are
separated by less than 3\arcsec, the largest X-ray/optical offset
being 2.9\arcsec. The sources are detected in 58 distinct fields.
The area covered for this sample is $\sim 5.8 \ {\rm deg}^2$ to
a flux limit of $\sim 10^{-15.2}$~\funits.
For each source we calculated $L_X$ by using $z$ and $f_X$.  We used
the value of $f_X$ provided by XASSIST.

We calculated HRs using the original ACIS event files and
the soft (hard) energy band $0.3-2.0$~keV ($2.0-8.0$~keV).

As in the previous Section, the sources were also checked visually.
The probability of detecting an optical counterpart
by chance is, once more, less than 1\%.

Details of our \chandra-\2df\ correlation sample are given in
Table~\ref{tab:chandra}.

\section{Classification of sources}\label{sec:sel}
In what follows we refer to the \xmm\ and \chandra\ correlation
samples together as \lq the \2df\ correlation sample\rq.  We now
describe our approach for identifying NGs in this 
sample.

For optical spectra it has been shown that empirical emission-line
intensity ratios may be used as a diagnostic of AGN activity
\citep[BPT diagrams,][]{1981PASP...93....5B}.  We adopted the
classification scheme of \citet{1997ApJS..112..315H} which is based on
the diagnostic diagrams proposed by
\citet{1987ApJS...63..295V}. \citet{1997ApJS..112..315H} used the line
ratios \othreef\ 5007/\hb, \oonef\ 6300/\ha, \ntwof\ 6583/\ha\ and
\stwof\ 6716, 6731/\ha, which are least sensitive to dust reddening
and flux calibration because the wavelength separation between members
of each line pair is small. Additionally, since the ratios involve a
line of only one element and an \hone\ Balmer line, they are less
abundance-sensitive.  We used this as our primary method of
classification.  Unfortunately, absolute flux calibration for \2df\
fibres, which can differ substantially in their throughput, cannot be
done reliably. For this reason we were unable to perform subtraction
of stellar templates which depends crucially on the shape of the
galaxy spectrum. As a result, we only classified galaxies as \lq
normal\rq, or otherwise, when all four of the line-intensity ratios
unambiguously suggested this. If only some lines were observable, the
classification is only tentative. This is denoted by a question mark
after the type entry in the sample tables.  Finally, some spectra were
tentatively classified as LINERS, according to the classification
criteria of \citet{1997ApJS..112..315H}.

In the case of sources 1, 2, 4 and 6 from the \chandra--\2df\
correlation sample the above method failed.
Using the tool PIMMS\footnote{http://heasarc.gsfc.nasa.gov/Tools/w3pimms.html}, we carried
out tests to estimate the hydrogen column density needed to
obtain the observed hardness ratio, assuming the source flux and
redshift, as well as a power law spectrum with photon index
around $\Gamma=1.8$. Our tests showed that for all sources \nh~$ >
10^{22}$, suggesting obscured AGN as indicated
in Table~\ref{tab:chandra}. However better optical observations are needed
to clarify the situation further, for the following reasons. For
source 1 the spectrum is of poor quality, whilst the \ha\ line is not
covered. For source 2, the \ha\ and \ntwof\ lines appear suppressed
but it is unclear whether this has a physical origin or is a data
reduction artefact. For source 4,
there is a hint of a broad \ha\ line, which only partially falls
within the \2df\ spectrum. In the case of source 6, the \ha\ line
appears absorbed and the \ntwof\ line falls outside the spectrum.

The emission-line ratio method also failed for sources 17 and 19.
However, as the optical spectra of these sources show
very broad \ha\ or \hb\ emission lines ($>1000$~\kmps),
the sources were also
classified as AGN, and this is indicated in Table~\ref{tab:chandra}.

Further, in order to separate NGs from AGN, it is often assumed that
for NGs, the X-ray--to--optical flux ratio obeys \lxo~$<-2$, i.e. is
two orders of magnitude lower than for typical AGN. The X-ray
luminosity is similarly assumed to obey $L_X \la
10^{42}$~\lunits~\citep{1989ARA&A..27...87F}.

For the \2df\ correlation sample we estimated \lxo\ from the relation
\begin{equation}\label{equ:stocke}
\log \frac{f_X}{f_O} = \log f_X({\rm 0.3-3.5 keV}) + 0.4V + 5.37
\end{equation}
\citep{1991ApJS...76..813S}. The X-ray flux for this relation was
calculated from the 0.2--8.0~keV (0.5--2.0~keV) flux of our XASSIST
(1XMM) dataset, assuming a power law index $\Gamma = 1.8$. The $V$
magnitude was calculated from the $b_j$ magnitude listed in the \2df\
database and used to select the objects optically. The conversion was
performed 
using the relations
\begin{equation}
b_j=B-0.28(B-V) \ ,
\end{equation}
which holds for the \2df\ galaxies and
\begin{equation}
B-V=0.53
\end{equation}
\citep[][Table 3a]{1995PASP..107..945F}. This is the average $B-V$
value for nearby galaxies with types in the range Sab, Sbc, Scd and
Irr.

We stress that we did not use either $L_X$ or \lxo\ to classify
sources. On the contrary, as explained above, where possible we first
classified sources as NGs or AGN using optical criteria. Subsequently, we used
$L_X$ to look for luminous NGs and checked whether the classification
agreed with the empirical AGN/NG break at \lxo$~\sim -2$.

\begin{table*} 
\small
\caption{NGs with \lxo~$>-2$. Groups of rows separated by horizontal
lines correspond to the sample indicated in the first column.  \2df\
stands for the combined correlation samples \2df--\xmm\ and
\2df--\chandra.  Each row corresponds to the NG type indicated in the
second column.  Each of the last three columns gives the fraction of
NGs with \lxo~$>-2$ in the $\log L_X$ region indicated at the top of
the column. Note that the second of these columns includes NGs from 
the $\log L_X < 42$ column which {\it also} have $\log L_X > 41$. 
Thus the numerator gives the number of NGs of a given type
in a given sample and a given $\log L_X$ range, and the denominator
the total number of NGs in the same sample and $\log L_X$ range.
Galaxies for which only upper limit information is available have not
been taken into account. Empty entries indicate that no NGs of this
type have been found. Galaxy labels are as in tables~\ref{tab:chandra}
and \ref{tab:xmm}. Additionally, sb stands for starburst and E for
elliptical.}
\label{tab:results} 
\centering 
\begin{tabular}{llc cc} 
\hline\hline 
Sample    & NG type & $ \log L_X < 42 $ & $41 < \log L_X < 42$ & $\log L_X \ge 42$ \\
\hline
\2df\     &\htwo\   &        4/18       &                      &          \\
          &\htwo?   &        1/18       &                      &          \\
          & A       &        2/18       &  2/2                 &   1/1   \\
\hline
\citet{ZezasPhD} & any & 0 & 0 & 0\\

\hline
\citet{1992ApJS...80..531F}&  sb & 2/167  &  1/26 & \\
                           &   E &  11/167  &  11/26 &    3/3\\
\hline






\end{tabular} 
\end{table*}

\begin{figure*}
\centering
\includegraphics[width=10cm,angle=-90]{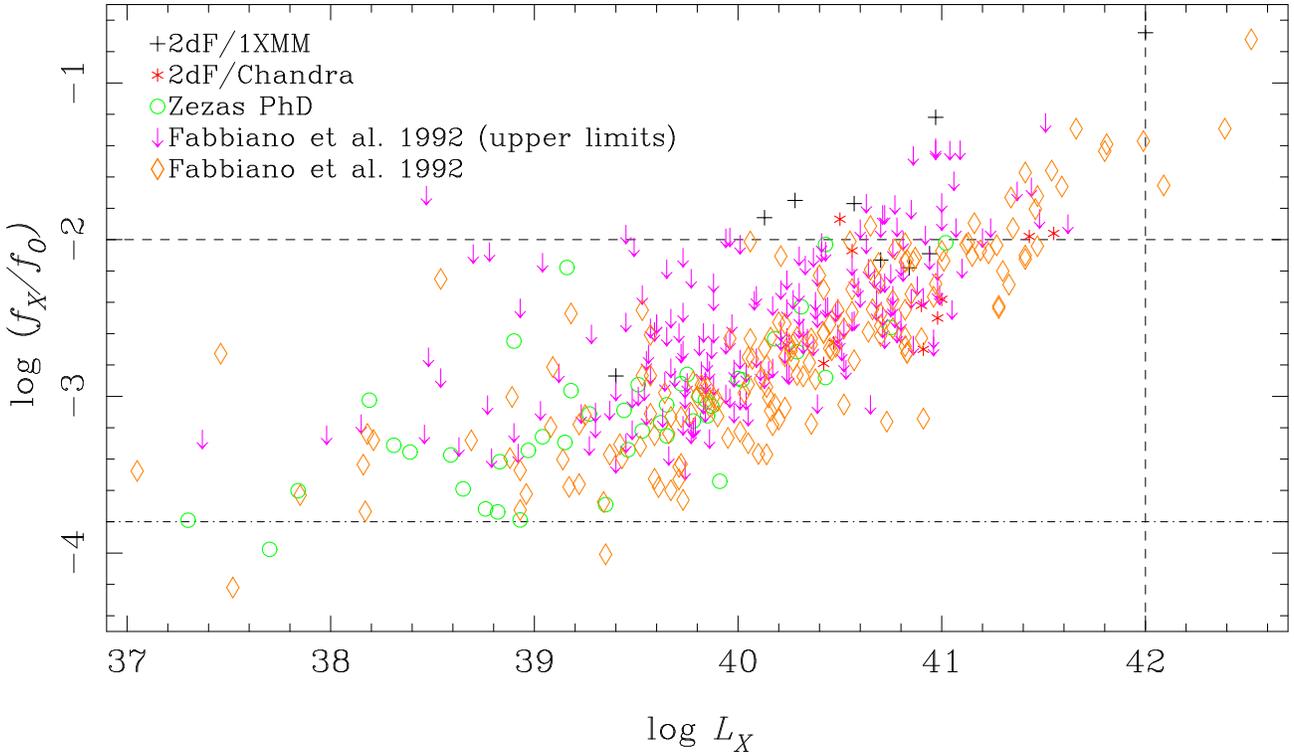}
\caption{Plot of \lxo\ versus $\log L_X$. Plotted here are values
calculated from our \xmm\ and \chandra\ data, as well as from the
literature (see legend in the plot).  For all samples only galaxies
which are classified as \htwo\ or absorption-line are plotted.  
All luminosities are for the same energy band ($0.5 - 2.0$ keV). The
two dashed lines demarcate the regions of \lxo$-L_X$ space which may
be inhabited mainly either by NGs ($\log L_X<42$, \lxo~$<-2$) or by
AGN. The dash-dotted line indicates an estimate for
\lxo\ from low mass X-ray binaries in a $10^{11} M_{\sun}$ 
galaxy (see text).}
\label{fig:fxfol}
\end{figure*}

\section{Results and discussion}
Results for the \2df\ correlation sample are summarised in
Tables~\ref{tab:xmm} and \ref{tab:chandra}. We found six NGs, which
fall under the category of \htwo\ nuclei, and two tentative \htwo\
nuclei.  Five galaxies are classified as AGN, and three as tentative
AGN.  There are eleven absorption line, presumably elliptical,
galaxies, for which the $L_X$, \lxo\ and HR values, considered
together, suggest that these are also not AGN-dominated. A further
three sources are tentatively classified as LINERs. For the rest of
the galaxies there is not enough information to make even a tentative
classification.  As can be seen in Tables~\ref{tab:xmm} and
\ref{tab:chandra} all \lq non-A\rq\ NGs have $\log L_X < 42$. In that
sense, no luminous star-forming galaxies have been found. However,
four of the \htwo\ nuclei, one of the tentative \htwo\ nuclei, as well
as three absorption-line galaxies have \lxo~$> -2$
(Table~\ref{tab:results}), i.e., overall, $\sim 40\%$ of NGs found
appear to have \lxo~$> -2$.

A plot of \lxo\ versus $L_X$ is shown in Fig.~\ref{fig:fxfol}.  Apart
from our \xmm\ and \chandra\ results, for comparison and completeness
purposes we have plotted data from the following sources:
\begin{enumerate}
\item The nearby star-forming galaxy sample compiled by \citet{ZezasPhD}. 
This comprises 44 galaxies detected by {\it ROSAT} PSPC, spanning the 
luminosity range $L_X({\rm 0.1 - 2.4 \ keV}) \approx 4\times 10^{37} -
3\times 10^{41}$\lunits. Galaxies in this sample have been
classified on the basis of high quality nuclear spectra from
\citet{1997ApJS..112..315H}.
\item The nearby galaxy sample of \citet{1992ApJS...80..531F}. The
galaxies have been observed with the {\it Einstein} observatory and
comprise all morphological types.  Galaxies flagged as AGN hosts in
the original sample have been excluded. We have also carried out a
further literature search to exclude more AGN from the final plotted
sample.  Galaxies for which only upper limit $f_X$ and $L_X$
information is available are plotted with downward-pointing arrows.
We used the $L_X$ values from \citet{1992ApJS...80..531F}, after
scaling for $H_0=72$~\hunits.  The \lxo\ values plotted were computed
as explained in Section~\ref{sec:sel}, using the $0.2 - 4.0$~keV flux
and $B$ magnitude information from \citet{1992ApJS...80..531F}, and
$B-V=0.655$. The latter is the average $B-V$ value for nearby galaxies
of all morphological types \citep[][Table 3a]{1995PASP..107..945F}.
\end{enumerate}

There is a clear correlation between the quantities plotted in the
figure. From the relation $L_X \sim L_B^{1.8}$
\citep{1992ApJS...80..531F} we would indeed expect a correlation of
the form, roughly, \lxo~$\sim 0.5 \log L_X$ as is the case in this
plot.  

The observed correlation shows that \lxo\ reaches values significantly
higher than $-2$ as $L_X$ becomes higher than $\log L_X \approx
41$. Thus, any survey for NGs which uses a \lxol2\ cut is likely to
suffer from incompleteness at the brightest luminosities.  Although
none of our samples can be said to be statistically complete, we may
assume that galaxies from different populations are randomly selected.
To assess the incompleteness we give in
Table~\ref{tab:results} the fraction of NGs in three $L_X$ regions per
NG type and sample.  Only NGs with \lxo~$>-2$ are included.  Galaxies
for which only upper limit values are available are not used.

From Fig.~\ref{fig:fxfol} and Table~\ref{tab:results} it is clear that all galaxies in the
sample of \citet{ZezasPhD} have both $\log L_X < 42$ and
\lxo~$<-2$. However, in the sample of \citet{1992ApJS...80..531F}
there is a number of galaxies for which \lxo~$>-2$.

\begin{figure}
\centering
\includegraphics[width=9cm,angle=-90]{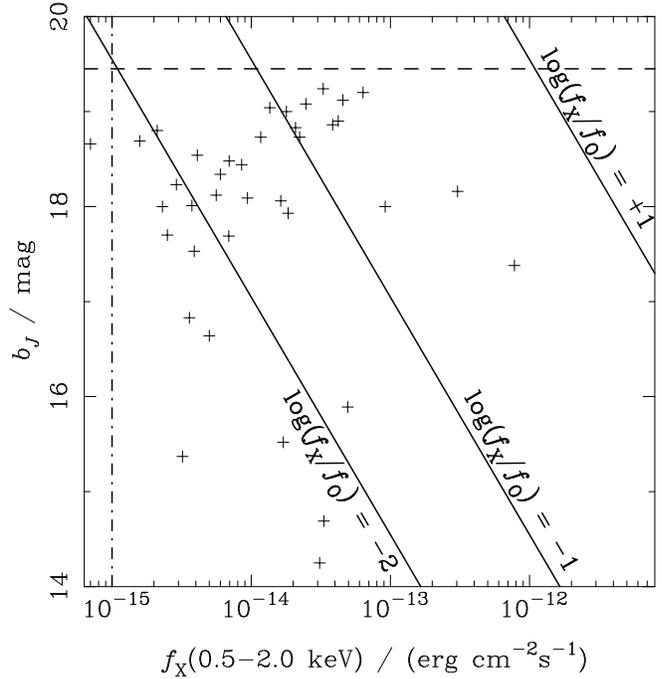}
\caption{Plot of $b_J$ magnitude versus logarithmic X-ray flux for all
galaxies in Tables~\ref{tab:xmm} and \ref{tab:chandra}. The region to
the right of the line \lxo~$=-1$ is expected to be occupied mainly by
AGN.  NGs are expected to be found mainly in the region to the left of
the line \lxo~$=-2$.}
\label{fig:bjfx}
\end{figure}

Specifically, as can be seen from Table~\ref{tab:results}, there are 13
galaxies from this sample, for which \lxo~$>-2$ and $\log L_X < 42$. All but one
of these galaxies have $41 < \log L_X < 42$. About 15\% are starbursts
(2/13) and the rest massive ellipticals, brightest in a group (BGGs)
or cluster (BCGs).
In the $\log L_X \ge 42$ region, there are 3 galaxies with
\lxo~$>-2$.
Two of these are massive elliptical BGGs and the third is
a cD BCG.

It is instructive to investigate how many galaxies with $\log L_X >
42$ we should expect to detect, given our survey's flux limit and 
area covered.  We used our flux limits and $\log L_X > 42$ to
integrate the local X-ray luminosity function of NGs
\citep{2005MNRAS.360..782G}. We found that we would expect to see $\la
1$ galaxy with each telescope.  In the \2df\ correlation sample there
is a single absorption line galaxy at $\log L_X = 42$, which is a good
NG candidate. There is thus good order of magnitude agreement between
our simple estimate and actual results.

It is also useful to compare observed values of \lxo\ to the numbers
expected due to low mass X-ray binaries (LXRBs), whose combined
luminosity has been shown to scale with the stellar mass of the host
galaxy \citep{2004MNRAS.349..146G}. We estimated the value of \lxo\ 
expected for
a $10^{11} M_{\sun}$ galaxy, using the $L_X ({\rm LXRB}) - M_{\ast}$
and $M_{\ast}/L_K - (B-V)$ correlations from \citet{2004MNRAS.349..146G}
(Figure 14 and Equation 2, respectively). We used the average value
of $B-V$ for all galaxies in our paper, and the average value of
$V-K$ for all galaxy types from \citet{2001MNRAS.326..745M}. We obtained 
the value \lxo$_{\rm LXRB} = - 3.8$, shown by the dot-dashed line
in Figure~\ref{fig:fxfol}. The bulk of the points in this plot
fall above this line, suggesting that this value is a good estimate
for a lower \lxo\ estimate.

Our sample is by no means complete.
In Figure~\ref{fig:bjfx} we plot $b_j$ magnitude against $0.5 - 2.0$
keV flux for all galaxies in the \2df\ correlation sample. The solid
oblique lines show $f_X-b_j$ loci for different values of \lxo. Our
galaxies are split in approximately equal numbers among the regions
demarcated by the constant \lxo\ lines. The dashed horizontal line
shows the \2df\ magnitude limit at $b_j=19.45$. The dashed-dotted
vertical line shows the approximate flux limit of our correlation
sample. It is clear that, to this limit, our survey misses galaxies
in the region $-2 <$~\lxo~$<-1$, where luminous NGs are likely to be
found. Such galaxies would not be missed only at a higher flux limit,
$\sim 2 \times 10^{-14}$~\funits. Furthermore,  
for galaxies brighter than $b_j \sim 16$,
saturation effects affect the completeness of \2df \citep{2002MNRAS.336..907N}.

The \citet{1992ApJS...80..531F} sample also suffers from increasing
incompleteness as $\log L_X$ increases. Table~\ref{tab:results} shows
that the fraction of NGs with \lxo~$>-2$ rises from $\sim8\%$ (13 out
of 167 galaxies) below $\log L_X=42$ to 100\% above $\log L_X=42$ (3
out of 3 galaxies). For $\log L_X < 42$ the majority of NGs with
\lxo~$>-2$ are massive ellipticals and for $\log L_X \ge 42$ all are
of this type. Although the \2df\ correlation sample has fewer
galaxies, it is clear that, at least at the highest luminosities,
there is a similar pattern, with absorption line galaxies making up
the bulk of NGs with \lxo~$>-2$. We note, however, that, considering
all $L_X$ regions together, the fraction of NGs
with \lxo~$>-2$ in the \2df\ correlation sample is much higher than in
the \citet{1992ApJS...80..531F} sample ($\sim 40\%$ versus $\sim
9\%$). This may suggest that the \2df\ correlation sample may suffer
from residual AGN contamination.  Furthermore, as mentioned in
Section~\ref{sec:sel}, this sample suffers from incompleteness at
bright optical magnitudes.

Considering all samples, a picture is thus emerging in which the \lxol2\
criterion selects against the brightest ellipticals, but is not
inadequate for other morphological types.  However, the ellipticals in
question, although \lq normal\rq\ in the sense that they do not host
an AGN, belong to a distinct sub-class with respect to star-forming
galaxies, as well as other ellipticals: these galaxies are found in
the centres of X-ray bright groups or clusters. Such systems are
affected significantly by their environment. For instance, they are,
on average, considerably more luminous than normal
ellipticals. Furthermore, group dominant galaxies have been shown to
have temperature profiles indicative of central cooling
\citep{2000MNRAS.315..356H}, leading to the suggestion that their
halos are actually the product of cooling flows associated with the
surrounding group.

The significance of the
findings from the \2df\ correlation samples is unclear, given the
uncertainty in morphological type and diagnostic emission line
ratios. Furthermore, classification of \2df\ galaxies which remain
unclassified in this work would make the current picture more
clear. However, flux calibrated spectra of higher signal-to-noise
ratio over the full wavelength range, covering all diagnostic emission
lines, are necessary for this to be achieved.

From a different perspective, we note that we found no confirmed AGN
with \lxol2. Regardless of any completeness problems, this shows that
the NHS surveys \citep{2004MNRAS.354..123G,2005MNRAS.360..782G} are at
least not significantly contaminated by AGN.

\section{Summary}  
We compiled a sample of galaxies observed with \xmm\ and \chandra\ for
which there are optical spectral observations in the \2df. We looked
for X-ray luminous ($L_X > 10^{42}$~\lunits) NGs and assessed the
empirical criterion \lxol2\ for separating NGs from AGN.  Our main
results are:
\begin{enumerate}
\item No luminous NGs were found.
\item Five \htwo\ galaxies and three absorption-line galaxies have
\lxo~$>-2$. 
\end{enumerate}

We also carried out the same type of search in two samples from the
literature. In the \citet{ZezasPhD} sample all galaxies have $\log L_X
< 42$ and \lxol2. In the \citet{1992ApJS...80..531F} sample, there are
two starbursts and fourteen ellipticals with \lxo~$>-2$. This
translates to an incompleteness of $\sim 8\%$ (100\%) for $\log L_X < 42$
($\log L_X \ge 42$).

Considering all samples, we thus find that the \lxol2\ criterion
seems to select primarily against the brightest, massive
ellipticals (BCGs and BGGs).

Further, we also find that the great majority of
our galaxies have \lxo~$> -3.8$ which represents an estimate
for the contribution of LXRBs to the X-ray luminosity of
galaxies.

\section{Acknowledgments}
We thank the anonymous referee for constructive comments which
helped improve the manuscript.
This work is funded in part by the Greek National Secretariat for
Research and Technology within the framework of the Greece-USA
collaboration programme {\it Study of Galaxies with the Chandra X-ray
Satellite}.  We acknowledge the use of data from the \xmm\ Science
Archive at VILSPA, the \chandra-XAssist archive and the \2df. This
research has made use of data obtained from the High Energy
Astrophysics Science Archive Research Center (HEASARC), provided by
NASA's Goddard Space Flight Center. This research has made use of the
NASA/IPAC Extragalactic Database (NED) which is operated by the Jet
Propulsion Laboratory, California Institute of Technology, under
contract with the National Aeronautics and Space Administration.

\end{document}